**Optical Push Broom On a Chip**

*Mahmoud A. Gaafar[1,2,*], He Li[3], Xinlun Cai[3], Juntao Li[3, **], Manfred Eich[1,4], and Alexander Yu. Petrov[1,5]*

[1]Institute of Optical and Electronic Materials, Hamburg University of Technology, Hamburg 21073, Germany
[2]Department of Physics, Faculty of Science, Menoufia University, Menoufia, Egypt
[3]State Key Laboratory of Optoelectronic Materials & Technology, Sun Yat-sen University, Guangzhou 510275, China
[4]Institute of Materials Research, Helmholtz-Zentrum Geesthacht, Max-Planck-Strasse 1, Geesthacht, D-21502, Germany
[5]ITMO University, 49 Kronverkskii Ave., 197101, St. Petersburg, Russia
[*]Corresponding author: mahmoud.gaafar@tuhh.de
[**]Corresponding author: lijt3@mail.sysu.edu.cn

**Abstract**

Here we report the first experimental demonstration of light trapping by a refractive index front in a silicon waveguide, the optical push broom effect. The front generated by a fast pump pulse collects and traps the energy of a CW signal with smaller group velocity and tuned near to the band gap of the Bragg grating introduced in the waveguide. This situation represents an optical analogue of light trapping in a tapered plasmonic waveguide where light is stopped without reflection. The energy of the CW signal is accumulated inside the front and distributed in frequency. In this experiment a 2 ps free carrier front was generated via two photon absorption of the pump in silicon waveguide. It collects approximately a 30 ps long packet of the CW signal. The presented effect can be utilized to compress signals in time and space.

## 1. Introduction

Dynamic frequency and bandwidth change of light is a highly desirable functionality in optical information processing [1, 2]. Particularly interesting is the phenomenon of light trapping by nonlinear pulse as this trapping increases the interaction time and thus the potentially achievable level of spatial/temporal compression. One of the prominent trapping mechanisms used in supercontinuum generation is provided by a decelerating soliton pulse, where a faster signal pulse coming from behind the soliton is trapped after interacting with the soliton [3–5]. In this case a signal that is reflected by trailing edge of the soliton, is trapped and, similar to optical analogue of event horizon [6, 7], cannot leave the soliton, which is continuously slowing down. The pulse trapping in this case can be explained by an effective 'gravity-like' potential that is produced by the decelerating soliton [4].

On the other hand, the trapping can also be induced by a refractive index front moving with a constant velocity in a waveguide with hyperbolic dispersion [8, 9]. The light interaction with a front leads to an indirect transition of light where its frequency and wavenumber both are changed [10–15]. The indirect transition normally occurs to the state in the unperturbed waveguide or states of perturbed waveguide, leading to the reflection from [16, 17] or transmission through the index front [11, 12, 18]. However, in the special case of hyperbolic dispersion a slow light signal can be trapped inside a fast co-propagating front if it does not find any state in the waveguide before or after the front after the interaction, i.e. does not undergo either inter- or intraband transitions. Alternatively speaking the signal is accelerated up to the front velocity and further interaction does not lead to group velocity change. This effect, named as optical "push broom", has been theoretically proposed by de Sterke [8, 19] and experimentally realized [9] in a fiber Bragg grating. Such trapping also leads to the pulse compression as the energy of the input signal is concentrated in the front.

In the last years different approaches of optical pulse temporal compression has been theoretically discussed and experimentally demonstrated. Electro-optic modulation [20] or four wave mixing (FWM) [21] induced time lenses were used to build a temporal telescope with compression factors up to 27 [21]. Such time lenses require fast electro-optic modulation, specially prepared pump pulses and adjusted dispersion compensation elements.

On the other hand, pulse compression can be obtained through front-induced transitions (FITs) via trapping [9, 14]. In case of trapping the accumulated frequency shift of the signal $\Delta\omega$ is proportional to the time spend inside the front $\Delta t$ and front slope [14]:

$$\Delta\omega = \left(\frac{\partial \Delta\omega_D}{\partial t_f}\right)\Delta t \tag{1}$$

where $t_f$ is the front duration and $\Delta\omega_D$ is the shift of the dispersion curve induced by the front. Therefore, different temporal parts of the signal automatically accumulate a linear frequency shift according to the time when they enter the front. This way, the signal time function is directly converted to frequency function, where the frequency distribution has a width equal to $(\Delta\omega_D/t_f)\tau$, where $\tau$ is the signal pulse duration. As an example, for the dispersion shift $\Delta\omega_D$ corresponding to 1 nm wavelength shift, front duration of $t_f = 1$ ps and a signal duration of 30 ps the bandwidth can approach 30 nm. This corresponds to pulse duration of 0.1 ps and the compression factor of 300.

However, in the push broom experiment by Broderick et. al. [9], due to the low Kerr nonlinearity of the fiber a very weak Bragg grating with band gap opening of 0.04 nm and an overall length of 8 cm had to be used. The frequency shift induced by the moving refractive index front in this experiment was not discussed, but it can be estimated from the duration of the signal being compressed from $\approx 300$ ps to 70 ps in the front to be in the order of 0.05 nm. Though a small portion of the CW signal was collected inside the front and the optical push broom effect was demonstrated, the investigation of the effect was not continued experimentally, e.g. no attempt was reported so far to use waveguide with higher nonlinearity and reduce the required propagation length. Compared to the fiber based approach integration of the push broom effect to sub-milimeter lengths on a silicon chip has many advantages. Refractive index perturbations in the order of 0.003 can be obtained in silicon with pulsed lasers having several picosecond pulse duration and peak power in the order of 10 W [11, 12, 17], compared to $10^6$ W required in fibers [9]. This low power induced perturbation would allow the switching of a stronger (compared to that of the silica fibre case) Bragg grating with a band gap opening in 1 nm range. Another advantage of silicon Bragg grating, is that we can easily control (engineer) the dispersion relation by changing the geometry of the structure, in contrast to fiber Bragg gratings, which are formed by writing a periodic variation of refractive index along the core of the fiber using an intense UV laser [22].

Here, we show experimentally for the first time the signal trapping inside a moving refractive index front in a silicon waveguide. We present the explanation of the trapping effect as an indirect photonic transition and discuss the conditions for required band gap opening and nonlinearity. With that we also identify an analogy to light stopping in tapered plasmonic waveguides [23–25]. The front is generated inside a silicon Bragg grating waveguide (SBGW) by two photon absorption (TPA) of a 2 ps long pump pulse at 1.55 µm wavelength with a peak power of 15 W which induces a free carrier (FC) density of $\approx 9 \cdot 10^{17}$/cm³. The strong pump pulse tuned well away from the resonance of a Bragg grating traps wave packets of another co-propagating weak CW signal wave which is tuned near to the band gap of the SBG. We show here, that in case of the trapping configuration 20 times more energy from the CW signal is converted to new frequencies, compared to the cross phase modulation (XPM) configuration [26, 27] –in which the signal and pump/front are moving with the same group velocity.

## 2. Theory and approach

Interaction of light signal with an index front leads to an indirect photonic transition with a simultaneous change of its frequency $\Delta\omega$ and wavenumber $\Delta\beta$ [10, 14]. The direction of the indirect transition is defined by the phase continuity at the front, where $\Delta\omega/\Delta\beta$ is equal to the front velocity $v_f$ [11, 28]. Therefore, the angle of the indirect transition induced by the front is defined by the velocity of the front. **Figure 1(a)** shows a schematic representation of four different indirect transitions in a highly dispersive system with hyperbolic dispersion. We consider here an example of an index front due to the effect of FCs generation. Solid and dashed curves correspond to the dispersion relations of the system in front of and behind the index front, respectively. In this schematic, the group velocities of the front (indicated by the slope of the orange arrow) and of the signal are co-directed for all transitions. The grey line represents the phase continuity line with a slope equal to the group velocity of the front (we show only one line for clarity). The red and blue circles indicate the initial and final states of the signal wave, respectively.

Transition 1 corresponds to signal transmission through the front, i.e. the final state of the signal after interacting with the front is behind the front (i.e. inter-band transition) [11]. Transition 2 describes the situation when the signal pulse does not find state behind the front and therefore reflects from it in the forward direction (i.e. intraband transition) [17]. However, if the phase continuity line does not cut either the unperturbed (solid black) or the completely perturbed (dashed black) dispersion curves, the signal will be trapped inside the index front, as it will not find any state inside the waveguide before or after the front (transition 3). For the realisation of the optical push broom effect, the front-induced band diagram shift $\Delta\omega_{DFC}$ should be sufficient, otherwise the signal will transmit through the front (transition 4 in **Figure 1(a)**). For example, to push broom a signal located at the zero group velocity position, the necessary band diagram shift should be $\Delta\omega_{\mathrm{DFC}} > \Delta\omega_{\mathrm{PBG}}/2$, where $\Delta\omega_{\mathrm{PBG}}$ is the photonic band gap (PBG) opening. However this value can be reduced by locating the signal away from the zero group velocity position, as we do in our experiment and is confirmed by simulations.

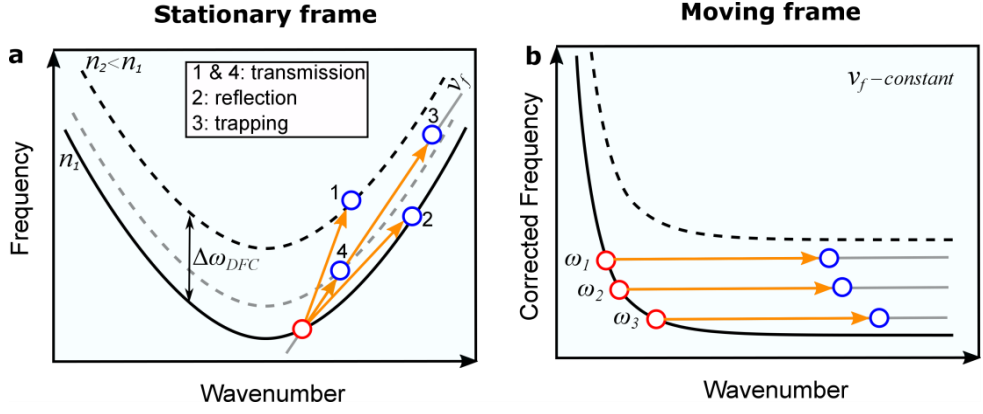

**Figure 1**: (a) Schematic representation in the stationary frame of different free-carrier-induced photonic transitions by changing the front velocity. Original band diagram is represented by the solid curve, while shifted band diagrams represented by the dashed black (strong index change) and dashed grey (weak index change). Red and blue circles indicate the initial and final state of signal wave, respectively. 1 and 4: transmission through the front, 2: reflection from the front and 3: signal trapping inside the front. (b) Schematic representation of the signal trapping (transition 3 in (a)) represented in the moving frame. When different frequency components ($\omega_1, \omega_2, \omega_3$) enter the index front they at the beginning will leave the original dispersion curve and will approach a constant positon between original and shifted dispersion curves, which corresponds to a certain position inside the front.

The transition can be also discussed in the frame moving with the front (**Figure 1(b)**) [14]. An interesting fact is, that in this frame, the corrected dispersion relation $\omega'(\beta) = \omega(\beta) - v_f \cdot (\beta - \beta_0)$ in case of trapping represents a saturating function with zero group velocity at infinite wavenumber $\beta$ (See **Figure 1(b)**). It reminds of the dispersion relation of a surface plasmon polariton [29]. When a surface plasmon polariton encounters a tapered perturbation that shifts the dispersion curve and does not allow further propagation then it is not reflected but is trapped at the transition leading to nanofocusing effects [23–25]. Different frequency components will freeze at different positions in the taper. The same is obtained now with a moving front in a waveguide with a hyperbolic dispersion curve $\omega(\beta)$. The signal enters the standing perturbation and stops inside the front without reflection. A monochromatic CW signal will approach zero group velocity in respect to the front and will stop at a defined location inside the front, thus ultimate compression would be achieved. In real situation an input signal has a finite duration and thus a certain frequency distribution in the corrected frame. Thus different spectral components will stop at slightly different locations inside the front.

The basic concept of front induced transition for trapping effect inside a SBGW is schematically shown in **Figure 2**. Such waveguide has a hyperbolic dispersion relation, which can appears also in periodic structures, such as Bragg stacks [30], fiber Bragg gratings [31, 32], silicon Bragg gratings [33, 34] and waveguides with periodical corrugation [35, 36]. We will conduct our experiment using a CW signal light and a pulsed pump similar to other work on pump probe interactions [6, 7, 9]. Using a CW signal will simplify the experimental verification of the frequency shift and will also facilitates the spectral filtering between the input signal and shifted signal.

In **Figure 2**, white color indicates the silicon material while gaps in between indicates air. Here, we consider a CW signal confined inside the waveguide and propagating with small group velocity. When a pump pulse is launched into the structure, it will generate FCs in the silicon waveguide and, consequently, induces a change of refractive index $\Delta n_{FC}$ which propagates with the velocity of the pump pulse $v_f$. The gradient of the induced index front corresponds to the pump pulse duration. In this schematic we consider only an index front due to the effect of FC generation and we neglect the Kerr effect. The white and red colors represent the waveguide before the pump pulse (with refractive index $n_1$) and after the propagation of the pump pulse (with reduced refractive index $n_2 < n_1$), respectively. The signal will be trapped at the front. As time progresses, the index front traps ever greater proportion of the signal, so that ultimately almost all of the initial signal energy is swept out of the waveguide.

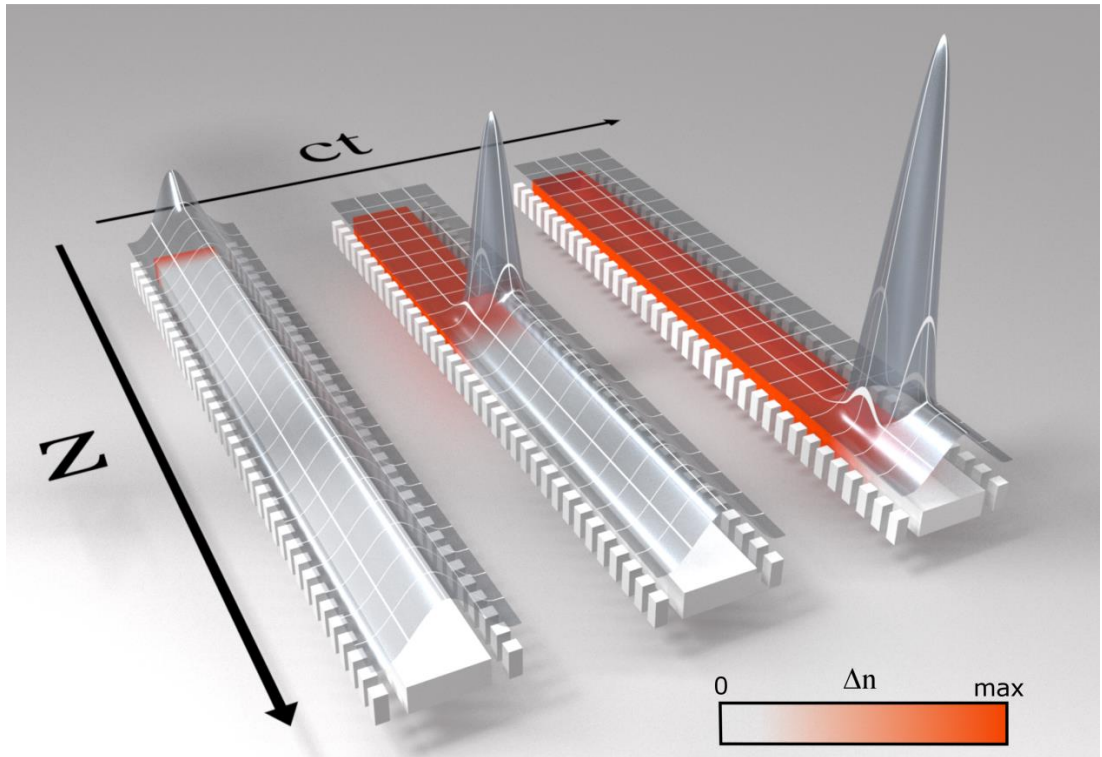

**Figure 2**: Schematic of the optical push broom effect inside a weak silicon Bragg grating waveguide. A refractive index front propagates with the velocity of the pump pulse (red colour). The fast front will frequency shift the slow signal wave and accelerate it to move with the front. The signal intensity is presented by a 3D surface, it accumulates at the front with time. The initial signal energy is swept out of the waveguide and no signal is present behind the front. In the presence of the free carries, the band shifts towards higher frequency and the CW signal that enters the waveguide after the front (pump) will be within the band gap, thus reflected at the input during the life time of free carriers.

**Figure 3** illustrates the advantage of the front induced light trapping to shift large portions of signal wave in one step, compared to the normal surfing/XPM. **Figure 3(a)** and **3(b)** represent a schematic of the spatial profile of the pump pulse at a given point of time and the corresponding spatial change of the waveguide's refractive index induced by the pump pulse, assuming ps pump pulses, respectively (see Supplemental Note 4) [26]. Typically, XPM leads to positive and negative frequency shift due to positive and negative slopes of the Kerr effect refractive index perturbation induced by the pump pulse in silicon [26, 27]. In this case, the time span of the signal wave shifted to a new frequency is limited the by pump pulse duration, as we can see in **Figure 3(c)**, where $t_{shifted}$ is related to $z_{shifted}$ as $t_{shifted} = \left( n_g^s \cdot z_{shifted}/c \right)$, where $n_g^s$ is the group velocity of the signal. Here, we consider a CW signal wave with constant power over time (solid grey line). Red and blue colors indicate the red and blue frequency shifted wave packets due to positive and negative slopes of the refractive index change, respectively (see Supplemental Note 4). We have to mention, that also in case of four wave mixing, the time span of the signal wave shifted to new frequency is limited by the pump pulse duration [37, 38]. However, in case of front induced signal trapping, the time span depends on both the group velocity differences between the pump $n_g^f$ and the signal $n_g^s$ and on the interaction length $L$:

$$t_{shifted} = L \left( \frac{\left| n_g^s - n_g^f \right|}{c} \right) \hspace{3cm} (2)$$

Thus, in this case we can envisage to shift large portions of CW signal in one step (see **Figure 3(d)**). We should mention, that **Figure 3(d)** is in the energy conserving approximation, thus the energy missing in the CW part is collected in the peak. However, in case of nonlinear losses the peak will reduce.

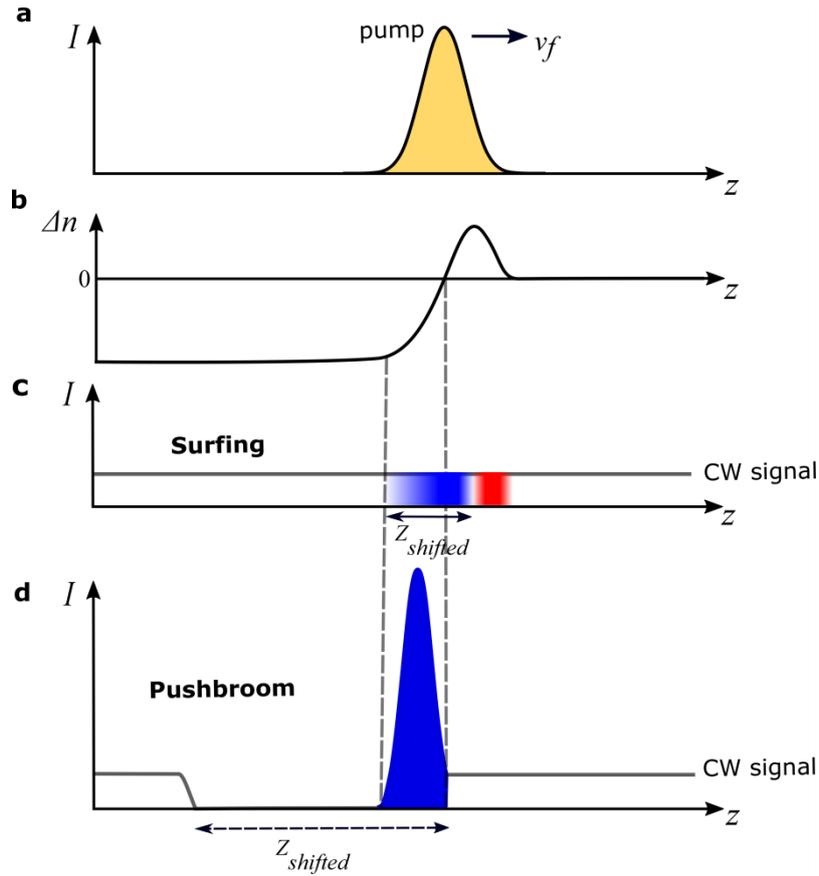

**Figure 3**: (a) Schematic of the spatial profile of the pump pulse at a given point of time, (b) the corresponding spatial change of the waveguide's refractive index induced by the pump pulse, due to both Kerr effect and FCs injections (see Supplemental Note 4), (c) and (d) Schematic representations of the CW signal wave (its power is represented by grey solid line) shifted to new frequency, in case of surfing and trapping, respectively. The length $z_{shifted}$ marks the length of the CW wave which is shifted to new frequencies. In case of surfing, the length of the shifted signal is limited by pump pulse size. Red and blue colors indicate the red and blue frequency shifted wave packets due to positive and negative slopes of the refractive index change, respectively. In case of front induced trapping, the shifted length can be much larger than pump pulse. Blue color again indicates the blue frequency shifted wave packets due the negative slope of the refractive index change. The red frequency shift is negligible in this case compared to the surfing case.

## 3. Experiments

### 3.1. Design and manufacture of silicon Bragg grating

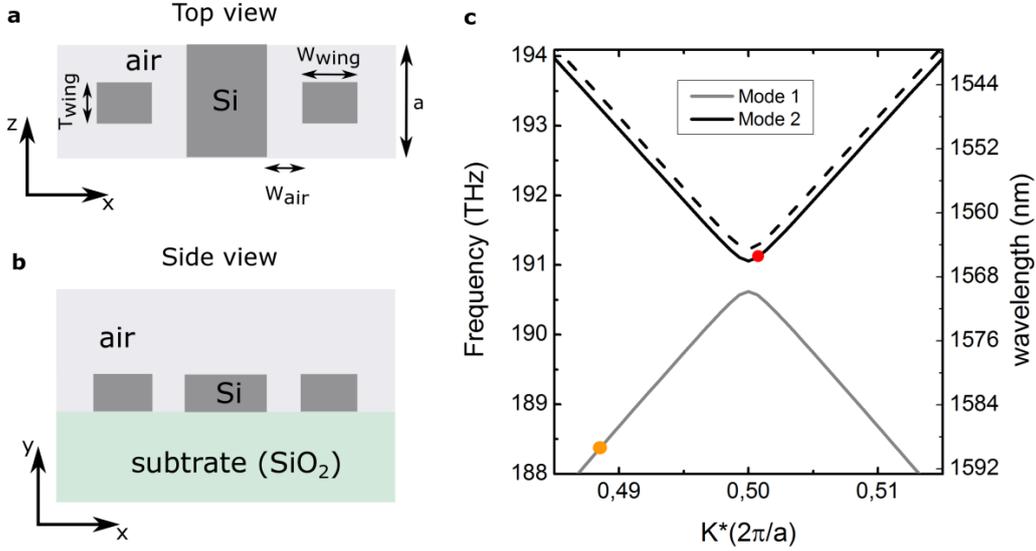

**Figure 4**: Schematic of the proposed system to obtain a weak silicon Bragg grating. (a) Top view of the Bragg grating. (b) Side view. The silicon slab height and width are 220 nm and 500 nm, respectively. The height, width ($W_{wing}$), and the thickness $T_{wing}$ of the silicon wings are 220 nm, 220 nm and 150 nm, respectively. The air gap between the silicon waveguide and the wings is $W_{air} = 40\ nm$, and the lattice constant a is 330 nm. (c) The simulated band diagram. Orange and red circles indicate the proposed pump and signal frequencies, respectively, for the experiment.

In order to implement the trapping effect on-chip, a silicon waveguide with ribs was designed similar to Refs. [35, 36]. A schematic of the waveguide design is shown in **Figure 4(a)** (top view of the structure) and **4(b)** (side view). Grey color indicates the silicon material, while white color indicates air. **Figure 4(c)** presents the simulated band diagram using CST microwave studio software. Solid black and grey lines represent the upper and lower branches of the dispersion relation, respectively, while the dashed black represents the shifted band due to the presence of FC front. Technically, if we positioned both the pump pulse and the signal wave on the upper branch of the dispersion relation, it will be challenging to detect the blue shifted signal due to overlap with the pump. Therefore, we position the pump frequency on the nondispersive part of the lower branch of the dispersion relation (orange circle in **Fig. 4(c)**).

The employed silicon waveguide has been fabricated on a Silicon-on-Insulator substrate with slab height of 220 nm and width of 500 nm. The height, width, and the thickness of the silicon wings (ribs) are 220 nm, 220 nm and 150 nm, respectively. The air gap between the silicon waveguide and the wings is $W_{air} = 40\ nm$, and the lattice constant a ≈ 330 nm. SEM image of the fabricated waveguide is shown in **Figure 5(a)**. Two grating couplers are used to couple the light into the Bragg grating. The grating coupler is 20 μm long including an adiabatic taper and 12 μm wide. The grating consists of air hole rows in silicon [39, 40]. This coupler has a 3 dB bandwidth of 30 nm with a peak coupling wavelength of 1570 nm, and a minimal coupling loss of ≈ 7 dB per coupler when the coupling angle is 13°. The detailed design and manufacturing method of the waveguide is given in Supplemental Note 2. The measured linear transmission (including coupling loss) and the group index of the TE-mode of the 1 mm silicon waveguide are shown in **Figure 5(b)** and **5(c)**, respectively. The shape of the transmission curve is due to the transmission characteristics of the grating coupler. For delay (group index) measurements, the transmission of a weak pulse with ≈ 0.4 nm bandwidth through the waveguide is measured by an optical sampling oscilloscope (which recovers the pulse durations from multiple repeating pulses). As we can see, the group index increases as the pulse approaches the band edge, i.e. the light slows down.

In our pump-probe experiments we consider two situations, cf. **Figure 3**. The first case is when the signal wave and index front, are moving with the same group velocities, i.e. the signal "surfing" on the front. The other case when the signal wave is excited at the band edge with group velocity smaller than the front velocity and is accelerated and trapped by the approaching pump pulse, then. Black and red dashed lines in **Figure 5(b)** indicate the wavelengths of the signal wave in case of surfing (1540 nm) and trapping (1564.8 nm), respectively. As seen from **Figure 5(c)**, the pump pulse at $\approx$1590 nm has a group velocity matched wavelength at 1540 nm ($n_g^f = n_g^{ss} = 3.5$), while the group index of the signal wave at wavelength of 1564.8 nm is $n_g^{st} \approx 12$. Far from the bandgap the dispersion relation of the silicon Bragg grating converges to that of the standard waveguide which still has some dispersion and there is small group velocity mismatch between signal at 1540 nm and pump at 1590 nm. But we neglect the effect of this mismatch in 1 mm waveguides that we use as accumulated time delay between two waves is approx. 0.3 ps which is much smaller than the pulse duration.

We have to mention also, that Kerr-induced self-phase-modulation (SPM) causes a temporally varying instantaneous frequency of the pump. This way, an initial non-chirped pump pulse acquires a frequency chirp. Still this frequency chirp does not change the pulse envelope, thus, the front shape as the pump is propagating at the frequency with negligible dispersion.

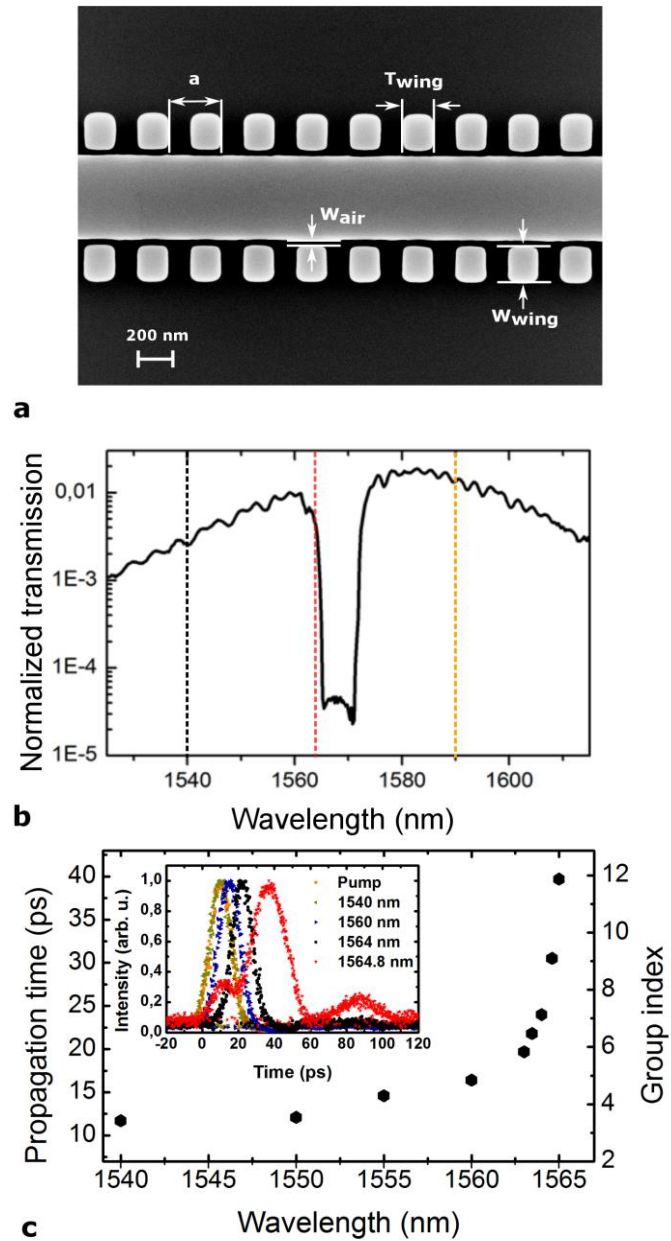

**Figure 5**: Characteristics of the fabricated silicon Bragg grating waveguide. (a) SEM of the device. (b) Linear transmission. Black and red dashed lines indicate the wavelengths of the signal wave in case of surfing (1540 nm) and trapping (1564.8 nm), respectively. The orange dashed line indicates the wavelength of the pump pulses. (c) The time delay and group index measurements. Inset shows the pulse temporal profiles, as the pulse approaches the band edge the third order dispersion distorts the pulse shape.

## 3.2. Signal surfing and trapping inside the front

We conduct our experiments using a CW signal light and a pulsed pump. Using CW simplifies the experimental verification of the frequency shift. A narrow CW spectrum facilitates spectral separation between the signal and shifted signal wavelengths. In this case, the portion of CW light interacting with the front can be determined from the group velocities of the pump and signal and the length of the silicon waveguide. This way, the energy at the shifted frequency can be directly compared to the energy in this portion of CW light resulting in the conversion efficiency. As we mentioned before, in our experiments we consider two cases: the signal wave "surfing" and the signal wave "trapping".

2 ps long pump pulses with 100 MHz repetition and 100 mW average power derived from mode-locked laser are launched into a 1 mm-long silicon Bragg grating waveguide at a center wavelength of 1590 nm with a group index of $n_g^f = 3.5$ (corresponding to orange dashed lines in **Figure 5(b)**). We also feed in the low power signal as a CW of light, which co-propagates with the index front in the waveguide. For the surfing case, the signal wavelength is 1540 nm with a group index of $n_g^{ss} = 3.5$ (corresponding to black dashed lines in **Figure 5(b)**), while for the trapping case, the signal wavelength is 1564.8 nm with a group index of $n_g^{st} = 12$ (corresponding to red dashed lines in **Figure 5(b)**). Details about the measurement setup are given in Supplemental Note 3.

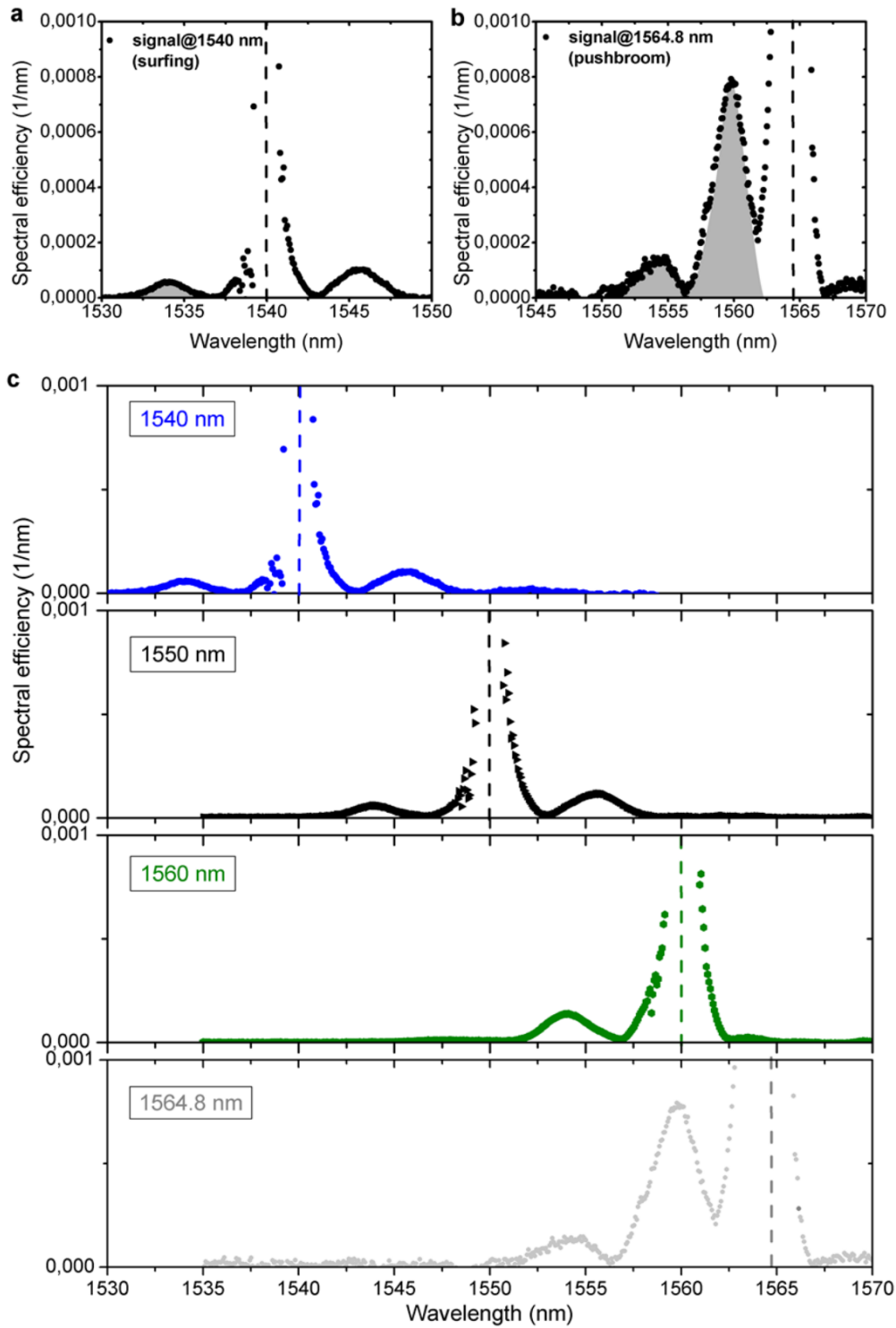

**Figure 6**: Experimental spectra recorded at the output of the 1 mm-long silicon Bragg grating waveguide in case of (a) surfing (signal wavelength at 1540 nm) and (b) trapping (signal wavelength at 1564.8 nm). Dashed lines represent the center wavelengths of the initial CW signals. Shown is the spectrum of the CW signal that has been frequency shifted. The spectral

efficiency is defined as the power per nm divided by the transmitted CW signal power. (c) Experimental output spectra for different input signal wavelengths.

**Figure 6** shows the experimental output spectra of the frequency shifted signal normalized by the output CW power of the transmitted signal for both surfing (**Figure 6(a)**) and trapping (**Figure 6(b)**) situations for input pump peak power of 15 W. The corresponding CW signal wave' powers at output were $\approx 4$ μW and $\approx 2$ μW, respectively. The spectral efficiency is defined as the power per wavelength divided by the power of the output CW signal. Traces with and without pump pulses present in the waveguide are recorded and subtracted, leading to the shifted signal displayed on a linear scale (see Supplemental Note 3). The center wavelengths of the initial signals are marked by dashed lines. As we can see, clear peaks appear on both blue and red sides of the initial input signal wavelengths in case of surfing, due to the positive and negative slope of the pump pulse. The trapping shows much larger conversion efficiency and much less red shifted signal (see also Supplemental Note 3). There is also large power density observed around original CW frequency. These are frequency components corresponding to the temporal 'shadow' left inside the CW signal after the cut-out of the wave packets interacting with the front and wave packets that cannot enter the Bragg grating while the free carriers are there.

**Figure 6(c)** represents the recorded spectra at the output of the waveguide for different input signal wavelengths at a fixed pump peak power. As we can see, as the signal approaches the band gap more energy is converted by the co-propagating index front.

## 4. Discussion

As we use a CW signal, only a small fraction of the total signal light can be converted within the finite length of the waveguide. The fraction η of the CW signal wave power that enters the front and is frequency converted including the FCs absorption is:

$$\eta = \eta_{CW} \cdot \eta_{loss} \qquad (3)$$

Where $\eta_{CW}$ is the fraction of the CW signal power that is frequency converted, given by:

$$\eta_{CW} = \nu_{rep} \cdot \left[ \left( \frac{n_g^s - n_g^f}{c} \right) \cdot L + t_p \right] \qquad (4)$$

Where $\nu_{rep}$ is the repetition rate of the pump pulses, $c$ is the velocity of light in vacuum, $L$ is the length of the waveguide and $t_p$ is the pump pulse duration. The part of the CW signal power that entered the front within the waveguide length $L$ is denoted by the first part in the bracket, while the second term relates to the CW signal wave that entered the waveguide when the pump pulse has just partly evolved for a fraction of its duration $t_p$.

While $\eta_{loss}$ is a factor taking into account the absorption of the signal wave packets by FCs during their interaction with the plasma front, and is given by:

$$\eta_{loss} = \frac{\int_0^L e^{-\alpha l} \cdot dl}{L} \qquad (5)$$

Where $\alpha$ is the FC absorption coefficient. For $\alpha = 30 \ cm^{-1}$ at FC concentration of $\approx 9.5 \cdot 10^{17}/cm^3$ [41] (see Supplementary Note 4) and L=1 mm, this leads to $\eta_{loss} \approx 30\%$.

We estimate the rise time of the negative front as $\tau = 1.9$ ps (see Supplemental note 4) which is comparable to the duration of the pump pulse. In case of surfing ($n_g^s = 3.5$), and for $\nu_{rep} = 100$ MHz, $n_g^f = 3.5$, and $L = 1$ mm, the maximal blue shifted fraction including the FC absorption of the incoming CW signal power due to the interaction with single pump pulses arriving at 100 MHz repetition rate is $0.57 \cdot 10^{-4}$. While in case of trapping ($n_g^s = 12$), the expected maximally transformable fraction is $0.95 \cdot 10^{-3}$. Thus trapping should be 17 times more efficient than in the surfing case.

From the measurements shown in **Figure 6(a) and (b)**, by integrating the area under blue shifted peaks for both surfing and trapping cases (marked by grey colors), we can see that the experimental blue-shifted conversion efficiency due to trapping is $\approx 0.71 \cdot 10^{-3}$ while due to surfing is $\approx 0.35 \cdot 10^{-4}$. Thus the experimental conversion efficiency due to trapping is approx.

20 times higher than in the surfing case, showing the great advantage of front induced transitions. In the measurement we obtain approximately 75% in case of trapping and 60% in case of surfing of what we expect from the theoretical estimation including the FCs absorption. The deviation can be attributed to inaccuracies in the estimation of the free carrier concentration.

It should be also mentioned that since a CW signal was used here, only $\eta_{CW}$=0.02-0.3% of the CW wave power was participating in the interaction process, and this value then represents the 100% conversion efficiency. As all power of the CW signal at times between the pump pulses not converted, thus, leading to a small apparent conversion efficiency and masking the effective conversion efficiency. When taking onto account only those fractions of the CW signal present at times when the interaction with the front takes place, the effective conversion efficiency of the signal wave packets trapped by the front is much larger and amounts to $\approx$ 25%. For signal pulses or pulse trains that have durations equal to the interaction time with the FCs front- instead of CW signal, we expect the conversion efficiency to approach this number.

In case of trapping the converted spectrum has a second peak at 1554 nm. It can be attributed to the CW wave packets which start initially inside the front, and therefore have the largest wavelength shift. We can estimate the maximum possible wavelength shift of the signal in the structure $\Delta\lambda = \Delta\lambda_D \cdot \Delta t/\tau$ , by assuming that the maximal time spent inside the front $\Delta t$ is equal to the travelling time of the signal inside the 1 mm structure which is $\approx$ 12 ps, $\Delta\lambda_D =$ 2.2 nm, and $\tau = 1.9$ ps (see Supplemental Note 4). This yields to a maximum expected possible shift of $\approx -13$ nm, which approximately fits to the maximal obtained experimental shift of $\approx -11$ nm.

From the observed bandwidth of the converted signal we can also estimate compressed pulse duration. The transform limited Gaussian pulse duration of a signal with 11 nm bandwidth is equal to $\approx 0.3$ ps. Thus, if other distortions can be neglected, a 30 ps part of converted CW signal is compressed 100 times.

We have to mention, that in the case of signal reflection from a copropagating index front reported before [17], the phase continuity line cuts through the unshifted band, thus the signal is reflected and stays in the waveguide before the front. We observe here, in case of push broom effect, 3 times larger CW conversion efficiency compared to the reflection effect [17] even though the pump in the push broom is fast and spends less time in the waveguide.

## 5. Conclusions

In conclusion, we have experimentally demonstrated a signal wave trapping inside a moving index front propagating in a 1mm long silicon Bragg grating waveguide. We have given an alternative explanation of the effect via an indirect photonic transition. With this explanation the effect presented here represents an optical analogue of light trapping in a tapered plasmonic waveguide, when a surface plasmon polariton is not reflected but stopped inside the waveguide. In the case of the front the light signal is not reflected from the front but is trapped inside. The trapping by a refractive index front in silicon waveguides also promises large frequency shifts and strong pulse compression. To demonstrate the effect we employed a CW signal and compared interactions in case of surfing, when signal and front have the same group velocities, and trapping, when slow signal light is caught by the fast front.

In contrast to surfing, the trapping configuration could collect 20 times more energy from the CW signal. A pulse compression in the order of 100 times is calculated from the frequency spectrum. With sharper fronts and longer waveguides compressions in the order of 1000 can be envisaged. This can be obtained, for example, if a 3 times longer waveguide is used, where input signal duration can be three times larger and final pulse is 3 times shorter due to a 3 times larger maximal frequency shift. This opens a way for efficient one step pulse compression in nonlinear waveguides. In addition, push broom effect can be used to realize a saturable-absorber-free pulsed laser in the mid-infrared wavelength from a CW laser. In this case large portions of the CW light will be cut out and compressed by the front without significant loss of average power.

## Acknowledgements

We would like to acknowledge the sponsorship from Dassault Systemes with their CST Studio Suite software. M. A. G, M. E, and A. Yu. P would like to acknowledge the German Research Foundation (DFG) (project number: 392102174) for its financial support. H. L, X. C, and J. L acknowledge the support from National Natural Science Foundation of China (NSFC) (11761131001).

M. A. G. performed the simulations, designed the waveguide and conducted the experiments, H. L., X. C. and J. L. fabricated the waveguide, J. L., M. E. and A. Yu. P. supervised the project, M. A. G, M. E. and A. Yu. P. analysed the results and wrote the paper.

## References

1   Almeida, V.R., Barrios, C.A., Panepucci, R.R., Lipson, M. (2004) All-optical control of light on a silicon chip. *Nature*, **431**, 1081.

2   Hau, L.V. (2008) Optical information processing in Bose–Einstein condensates. *Nature Photonics*, **2**, 451.

3   Nishizawa, N. and Goto, T. (2002) Pulse trapping by ultrashort soliton pulses in optical fibers across zero-dispersion wavelength. *Opt. Lett.*, **27** (3), 152–154.

4   Gorbach, A.V. and Skryabin, D.V. (2007) Light trapping in gravity-like potentials and expansion of supercontinuum spectra in photonic-crystal fibres. *Nature Photonics*, **1**, 653.

5   Hill, S., Kuklewicz, C.E., Leonhardt, U., König, F. (2009) Evolution of light trapped by a soliton in a microstructured fiber. *Opt. Express*, **17** (16), 13588–13601.

6   Philbin, T.G., Kuklewicz, C., Robertson, S., Hill, S., König, F., Leonhardt, U. (2008) Fiber-Optical Analog of the Event Horizon. *Science*, **319** (5868), 1367.

7   Ciret, C., Leo, F., Kuyken, B., Roelkens, G., Gorza, S.-P. (2016) Observation of an optical event horizon in a silicon-on-insulator photonic wire waveguide. *Opt. Express*, **24** (1), 114–124.

8   Sterke, C.M. de (1992) Optical push broom. *Opt. Lett.*, **17** (13), 914–916.

9   Broderick, N.G.R., Taverner, D., Richardson, D.J., Ibsen, M., Laming, R.I. (1997) Optical Pulse Compression in Fiber Bragg Gratings. *Phys. Rev. Lett.*, **79** (23), 4566–4569.

10  Yu, Z. and Fan, S. (2009) Complete optical isolation created by indirect interband photonic transitions. *Nat. Photon.*, **3** (2), 91–94.

11  Castellanos Muñoz, M., Petrov, A.Y., O'Faolain, L., Li, J., Krauss, T.F., Eich, M. (2014) Optically Induced Indirect Photonic Transitions in a Slow Light Photonic Crystal Waveguide. *Phys. Rev. Lett.*, **112** (5), 53904.

12  Kondo, K. and Baba, T. (2014) Dynamic Wavelength Conversion in Copropagating Slow-Light Pulses. *Phys. Rev. Lett.*, **112** (22), 223904.

13  Gaafar, M.A., Renner, H., Petrov, A.Y., Eich, M. (2019) Linear Schrödinger equation with temporal evolution for front induced transitions. *Opt. Express*, **27** (15), 21273–21284.

14  Gaafar, M.A., Baba, T., Eich, M., Petrov, A.Y. (2019) Front-induced transitions. *Nature Photonics*, **13** (11), 737–748.

15  Mahmoud Gaafar (2020) Pulse time reversal and stopping by a refractive index front. *APL Photonics*, **5**.

16  Plansinis, B.W., Donaldson, W.R., Agrawal, G.P. (2015) What is the Temporal Analog of Reflection and Refraction of Optical Beams? *Phys. Rev. Lett.*, **115** (18), 183901.

17  Gaafar, M.A., Jalas, D., O'Faolain, L., Li, J., Krauss, T.F., Petrov, A.Y., Eich, M. (2018) Reflection from a free carrier front via an intraband indirect photonic transition. *Nature Communications*, **9** (1), 1447.

18  Eilenberger, F., Kabakova, I.V., Sterke, C.M. de, Eggleton, B.J., Pertsch, T. (2013) Cavity Optical Pulse Extraction: ultra-short pulse generation as seeded Hawking radiation. *Scientific Reports*, **3**, 2607.

19  Steel, M.J. and Martijn de Sterke, C. (1994) Schrödinger equation description for cross-phase modulation in grating structures. *Phys. Rev. A*, **49** (6), 5048–5055.

20  Karpiński, M., Jachura, M., Wright, L.J., Smith, B.J. (2016) Bandwidth manipulation of quantum light by an electro-optic time lens. *Nature Photonics*, **11**, 53.

21  Foster, M.A., Salem, R., Okawachi, Y., Turner-Foster, A.C., Lipson, M., Gaeta, A.L. (2009) Ultrafast waveform compression using a time-domain telescope. *Nature Photonics*, **3**, 581.

22  Hill, K.O. and Meltz, G. (1997) Fiber Bragg grating technology fundamentals and overview. *Journal of Lightwave Technology*, **15** (8), 1263–1276.

23  Tsakmakidis, K.L., Boardman, A.D., Hess, O. (2007) 'Trapped rainbow' storage of light in metamaterials. *Nature*, **450** (7168), 397–401.

24  Gramotnev, D.K. and Bozhevolnyi, S.I. (2013) Nanofocusing of electromagnetic radiation. *Nature Photonics*, **8**, 13.

25  Stockman, M.I. (2004) Nanofocusing of Optical Energy in Tapered Plasmonic Waveguides. *Phys. Rev. Lett.*, **93** (13), 137404.

26  Dekker, R., Driessen, A., Wahlbrink, T., Moormann, C., Niehusmann, J., Först, M. (2006) Ultrafast Kerr-induced all-optical wavelength conversion in silicon waveguides using 1.55 μm femtosecond pulses. *Opt. Express*, **14** (18), 8336–8346.

27  Hsieh, I.-W., Chen, X., Dadap, J.I., Panoiu, N.C., Osgood, R.M., McNab, S.J., Vlasov, Y.A. (2007) Cross-phase modulation-induced spectral and temporal effects on co-propagating femtosecond pulses in silicon photonic wires. *Opt. Express*, **15** (3), 1135–1146.

28  Gaafar, M.A., Petrov, A.Y., Eich, M. (2017) Free Carrier Front Induced Indirect Photonic Transitions: A New Paradigm for Frequency Manipulation on Chip. *ACS Photonics*, **4** (11), 2751–2758.

29  Stefan Alexander Maier (2007) *Plasmonics: Fundamentals and Applications*, Springer US.

30  Pochi Yeh (2005) *Optical Waves in Layered Media*, Wiley.

31  Eggleton, B.J., Slusher, R.E., Sterke, C.M. de, Krug, P.A., Sipe, J.E. (1996) Bragg Grating Solitons. *Phys. Rev. Lett.*, **76** (10), 1627–1630.

32  Meltz, G., Morey, W.W., Glenn, W.H. (1989) Formation of Bragg gratings in optical fibers by a transverse holographic method. *Opt. Lett.*, **14** (15), 823–825.

33  Kaushal, S., Cheng, R., Ma, M., Mistry, A., Burla, M., Chrostowski, L., Azaña, J. (2018) Optical signal processing based on silicon photonics waveguide Bragg gratings: review. *Frontiers of Optoelectronics*, **11** (2), 163–188.

34  Sahin, E., Ooi, K.J.A., Png, C.E., Tan, D.T.H. (2017) Large, scalable dispersion engineering using cladding-modulated Bragg gratings on a silicon chip. *Appl. Phys. Lett.*, **110** (16), 161113.

35  Verbist, M., van Thourhout, D., Bogaerts, W. (2013) Weak gratings in silicon-on-insulator for spectral filters based on volume holography. *Opt. Lett.*, **38** (3), 386–388.

36  Verbist, M., Bogaerts, W., van Thourhout, D. (2014) Design of Weak 1-D Bragg Grating Filters in SOI Waveguides Using Volume Holography Techniques. *Journal of Lightwave Technology*, **32** (10), 1915–1920.

37  Monat, C., Corcoran, B., Pudo, D., Ebnali-Heidari, M., Grillet, C., Pelusi, M.D., Moss, D.J., Eggleton, B.J., White, T.P., O'Faolain, L., Krauss, T.F. (2010) Slow Light Enhanced

Nonlinear Optics in Silicon Photonic Crystal Waveguides. *IEEE Journal of Selected Topics in Quantum Electronics*, **16** (1), 344–356.

38 Monat, C., Ebnali-Heidari, M., Grillet, C., Corcoran, B., Eggleton, B.J., White, T.P., O'Faolain, L., Li, J., Krauss, T.F. (2010) Four-wave mixing in slow light engineered silicon photonic crystal waveguides. *Opt. Express*, **18** (22), 22915–22927.

39 Ding, Y., Ou, H., Peucheret, C. (2013) Ultrahigh-efficiency apodized grating coupler using fully etched photonic crystals. *Opt. Lett.*, **38** (15), 2732–2734.

40 Ding, Y., Peucheret, C., Ou, H., Yvind, K. (2014) Fully etched apodized grating coupler on the SOI platform with −0.58 dB coupling efficiency. *Opt. Lett.*, **39** (18), 5348–5350.

41 Schroder, D.K., Thomas, R.N., Swartz, J.C. (1978) Free Carrier Absorption in Silicon. *IEEE Journal of Solid-State Circuits*, **13** (1), 180–187.

42 Soref, R. and Bennett, B. (1987) Electrooptical effects in silicon. *IEEE J. Quant. Electron.*, **23** (1), 123–129.

43 J. D. Joannopoulos *Photonic Crystals, Molding the Flow of Light*, Princeton University Press, 17–19 (2008), Princeton, NJ.

44 Li, J., O'Faolain, L., Rey, I.H., Krauss, T.F. (2011) Four-wave mixing in photonic crystal waveguides: slow light enhancement and limitations. *Opt. Express*, **19** (5), 4458–4463.

# Supporting Information

**Optical Push Broom On a Chip**

*Mahmoud A. Gaafar, He Li, Xinlun Cai, Juntao Li, Manfred Eich, and Alexander Yu. Petrov*

Here we present the simulation details of the CW signal trapping inside the 1 mm long waveguide via a ray tracing approach. Furthermore, the experimental setup and measurement details are also shown here.

**Supplementary Note 1: Simulation of CW wave packets trajectories.** We simulate the CW signal trapping inside the 1 mm long waveguide via a ray tracing approach. We can split the CW signal into wave packets and track their trajectory. The approach discussed here describes propagation of a light ray in space and time in contrast to typical ray optics approximation in 2D or 3D space. Such calculation is applicable to any waveguide with known dispersion and not limited to low contrast waveguides. It is established for fiber optics [6] and photonic crystal waveguides [12]. This approach is limited only by the temporal size of the wavepacket, similar as it is limited by the ray width in case of spatial ray. Namely the wavepacket in time should be much larger than the period of oscillation. Furthermore, it neglects the diffraction and interference effects between the wavepackets. Still it helps to predict the position of the trapped energy inside the front and explain the pulse compression effect.

In our simulation we start with a defined frequency of the wave packet and calculate its wavenumber and group velocity at the input in accordance to its position in respect to the front. In a small time step a wave packet propagates a small distance according to its group velocity. Being inside the front the wave packet accumulates a frequency shift according to $\Delta\omega = \partial\Delta\omega_D/\partial t \cdot \Delta t$, a wavenumber shift according to $\Delta\beta = \Delta\omega/v_f$. This leads to a change of the group velocity which determines how the wave packet further propagates. Thus, we track the signal wave packets in space and time.

Here, we simulate the CW signal trapping inside the 1 mm long SBGW with parameters corresponding to experimental conditions. We assume that we can split the CW signal into wave packets located at different positions in the waveguide at a fixed point in time. We then launch these wave packets and follow their trajectories in a ray tracing approach.

We use light at 1.565 μm in a waveguide with a hyperbolic dispersion relation $\omega(\beta) = \omega_{PBG} + \Delta\omega_{PBG} \cdot \sqrt{1 + [(\beta - \beta_{PBG})^2/\Delta\beta_{PBG}^2]}$, emulating an upper branch of a dispersion relation with a PBG, employing a PBG half opening of $\Delta\omega_{PBG} = 2.198 \cdot 10^{12}$ 1/s ($\Delta f_{PBG} = 0.35$ THz ) and a PBG center frequency $\omega_{PBG} = 1.2 \cdot 10^{15}$ 1/s ($f_{PBG} = 191.2\ THz$ ). $\Delta\beta_{PBG} = \Delta\omega_{PBG}/v_{g\infty}$ is the parameter that is chosen in such a way, that away from the band edge the dispersion relation converges to a straight line with group velocity $v_{g\infty} = c/4.1$, as shown in **Figure S1(a)**, where $\omega_0 = \omega_{PBG} + \Delta\omega_{PBG}$. Here, $\beta_{PBG}$ is the center wavenumber of the PBG, and $c$ is the velocity of light in vacuum. The band diagram shift induced by the front corresponding to both Kerr effect and FCs injections is described by the function $\Delta\omega_D(t) = \Delta\omega_{DFC}/2 \cdot \left[1 + \text{erf}\left(\sqrt{4ln2}\ t/t_p\right)\right] - \Delta\omega_{DKerr} \cdot \exp\left[-4ln2t^2/t_p^2\right]$ , where $t_p = 2$ ps is the temporal pump pulse width, $\Delta\omega_{DFC} = 1.38 \cdot 10^{12}$ 1/s ($\Delta f_{DFC} = 0.22$ THz) and $\Delta\omega_{DKerr} = 0.69 \cdot 10^{12}$ 1/s ($\Delta f_{DKerr} = 0.11$ THz) are the maximum vertical band diagram shifts in frequency due to FCs injections and Kerr effect, respectively (see Supplemental Note 4). The input signal pulse is centered at frequency of 191.6 THz and has a velocity of $v_{g1} = c/14$, while the front has a velocity of $v_f = c/4.1$. The signal will be trapped and gradually frequency shifted along the phase continuity line (orange arrows in **Figure S1(a) and S1(c)** as long as it stays inside the front.

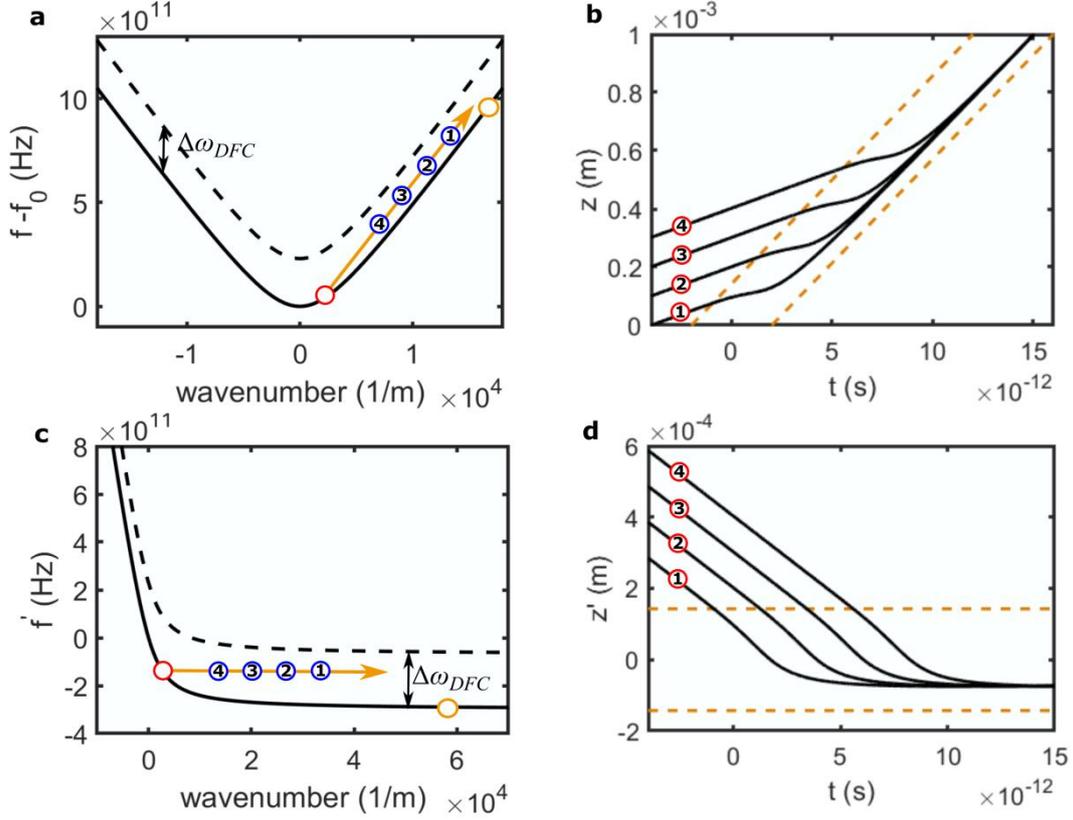

**Figure S1**: Simulation of optical push broom inside 1 mm long SBGW via ray tracing. (a) Original and shifted band diagram with a phase continuity line (orange line). The red and blue circles indicate the initial (slow mode) and final (fast mode) states of a signal wave, respectively. (b) The trajectories of four signal wave packets (black curves) represented in the stationary frame. The dashed orange lines represent the boundaries of $\pm t_p$. (c) The dispersion relation adapted for the moving frame. (d) The trajectories of the signal in the frame moving with the front. The dashed horizontal orange lines represent the boundaries of $\pm z_p$, where $z_p = v_f \cdot t_p$ is the spatial pump width.

The simulation results of the optical push broom effect in the stationary (laboratory) frame are presented in **Figure S1(b)**. We consider the CW signal wave as being composed of individual wave packets located at different z-coordinate positions at a given point in time (they are denoted by the numbers 1-4). These wave packets then interact with the front with a delay, depending on their relative positions to the front at a given time. Their trajectories are shown by solid black lines. The dashed orange lines represent the boundaries of $\pm t_p$. Here, we show the trajectories of only four wave packets for clarity. Before the faster approaching index front encounters the wave packet, this wave packet moves along a straight line with constant group velocity of the signal $v_{g1}$, which corresponds to the slope of the original dispersion curve at the original frequency. In the beginning of the interaction the signal is gradually decelerated, due to the Kerr-induced positive slope, and then gradually accelerated until its velocity reaches that of the front. As can be seen all wave packets of the same frequency converge to the same position inside the front.

The frequency change $\Delta\omega$ of the wave packet along the trajectory is proportional to the time spent inside the front $\Delta t$ (Equation 1) [28]. As different wave packets spend different time $\Delta t$ inside the front, they accumulate different frequency shifts, which leads to a broadening of the temporal and spatial frequency spectra (cf. several blue circles in **Figure. S1(a)** and **S1(c)**) and accordingly a narrowing in time and space of the output signal pulse, as shown in **Figure S1(b)**. Numbers 1,2,3,4 denote wave packets of the CW signal of which #4 is the one farthest ahead and #1 is the last. Thus, 1 interacts first with the pump pulse approaching from behind. Therefore, the maximal wavelength shift is accumulated by the wave packet #1, which spends the longest time inside the front, compared to the other wave packets.

**Figure S1(c)** shows the dispersion relation adapted for the moving frame $\omega'(\beta) = \omega(\beta) - v_f \cdot (\beta - \beta_0)$. In case of trapping it represents a saturating function with zero group velocity at infinite wavenumber $\beta$. In this corrected frame $z' = z - v_f t$, the signal enters the standing perturbation and stops inside the front without reflection (**Figure S1(d)**). As we launch here a monochromatic signal it stops at a defined location inside the front where its group velocity becomes zero, thus ultimate compression would be achieved.

**Supplementary Note 2: Structure of the silicon Bragg grating waveguide.** The employed silicon Bragg grating waveguide has been fabricated on a Silicon-on-Insulator substrate (SOI) wafer with a 220 nm thick silicon slab. The patterns were exposed on a spin-coated AR-P 6200.13 electron-beam photoresist by using the electron-beam lithography (EBL), which are then transferred into the silicon slab by using an inductively coupled plasma (ICP) dry etching. The residual photoresist is removed by the Oxygen plasma etching and a microresist remover. The Bragg grating waveguide consist of a straight waveguide with periodic islands (wings) on both sides. Two grating couplers are used to couple the light inside the Bragg grating. To obtain the suitable transmission gap, the grating waveguide has a width of 500 nm and the islands have the period of 324 nm. The height, width, and the thickness of the silicon wings are 220 nm, 220 nm and 190 nm, respectively, while the air gap between the islands and the straight waveguide is 35 nm. The straight waveguide is laterally tapered from 12 µm to 500 nm with 400 µm lengths at both ends to the grating couplers. The length of the waveguide with wings is 1 mm. A SEM image of the fabricated waveguide is shown in **Figure S2(a)**.

The fully etched grating couplers for efficient coupling between single mode fibers and silicon chips are fabricated in a single step of standard SOI processing, including EBL and ICP etching. The grating coupler is 20 µm long and 12 µm wide and consists of photoni crystal (PhC) columns with apodized hole sizes [39, 40]. This coupler has a 3 dB bandwidth of 30 nm with a peak coupling wavelength of 1570 nm, and a minimal coupling loss of 7 dB per coupler when the coupling angle is 13°. SEM of the grating coupler is shown in **Figure S2(b)**.

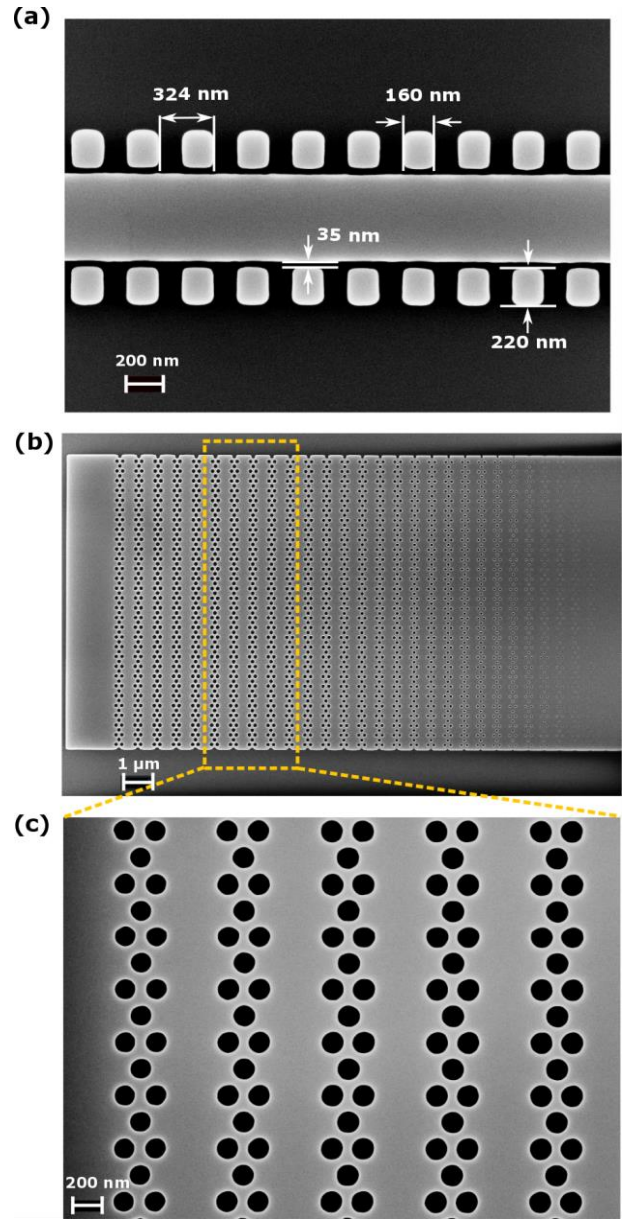

**Figure S2**. (a) SEM of the silicon Bragg grating waveguide. (b) SEM of the grating coupler.

**Supplementary Note 3: Measurement setup.** The experiment is arranged as displayed in **Figure S3**. A mode-locked fiber laser (Menlo system) with a 100 MHz repetition rate delivers pulses of approx. 100 fs duration with a wide spectrum ranging from 1500nm–1610nm. The output is connected to a grating filter (Filter 1) with adjustable bandwidth and center frequency followed by an erbium-doped fiber amplifier (EDFA). The output after EDFA is connected again to a second grating filter (Filter 2) to suppress the noise level at signal wavelength. Pulses with tunable center wavelength between 1525 nm –1600 nm and bandwidth between 0.1 nm – 4 nm (33 ps – 2 ps) can be produced from this configuration. The signal light is derived from a tunable diode laser (Photonetics PRI). It delivers up to 8 mW of CW light, tunable from 1500 to 1600 nm. The adjustment of optical polarization is carried out with a polarization controller. The pump pulse is subsequently combined with the CW signal light through a 50/50 (3 dB) beam combiner, which is then fiber-coupled to the Bragg grating using grating coupler. The optical spectra are measured by the optical spectrum analyzer.

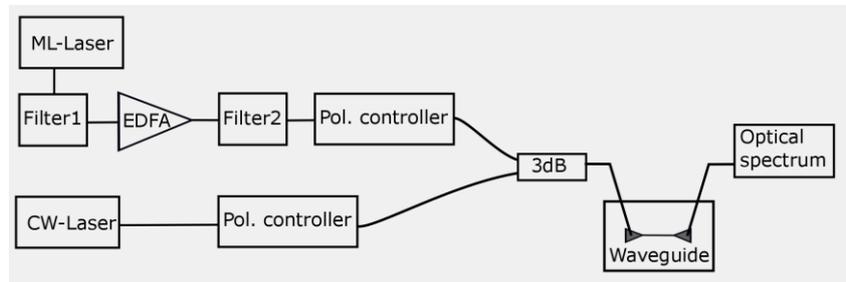

**Figure S3**: Schematic of the experimental setup. EDFA, erbium-doped fiber amplifier; Pol., polarizer and 3dB, 50:50 beam combiner. All solid lines represent fiber coupling.

**Figure S4** shows the experimental output spectra at the output of our silicon Bragg grating waveguide for a transmitted CW signal wave's power of (a) ≈ 4 μW at a wavelength of 1540 nm and (b) ≈ 2 μW at a wavelength of 1564.8 nm at a fixed pump peak power of 15 W. The spectrum of the CW signal that has not interacted with the pump pulses (red trace), of the pump pulses alone (orange trace), and of both the pump pulses and the signal (blue trace) are shown on a logarithmic scale in the upper graph. While, they are subtracted on a linear scale (black trace) and displayed in the lower graph. We performed the subtraction on a linear scale in order to obtain only the converted signal without the residual of the CW input signal, in contrast to subtraction in log scale which would give as a ratio. The dashed line represents the center wavelength of the input signals. The signal spectra in the absence of the pump have components outside of CW line due to noise amplification in the EDFA amplifier. As we can see, in the presence of the pump pulse, clear peaks appear on both blue and red sides of the initial input signal wavelength in case of surfing, due to the positive and negative slope of the pump pulse, while the red frequency shift is negligible in case of trapping compared to the surfing case. We have to mention, that the spectra of the signals without and with the pump is present have similar spectral features, which result from back reflections in the optical chip.

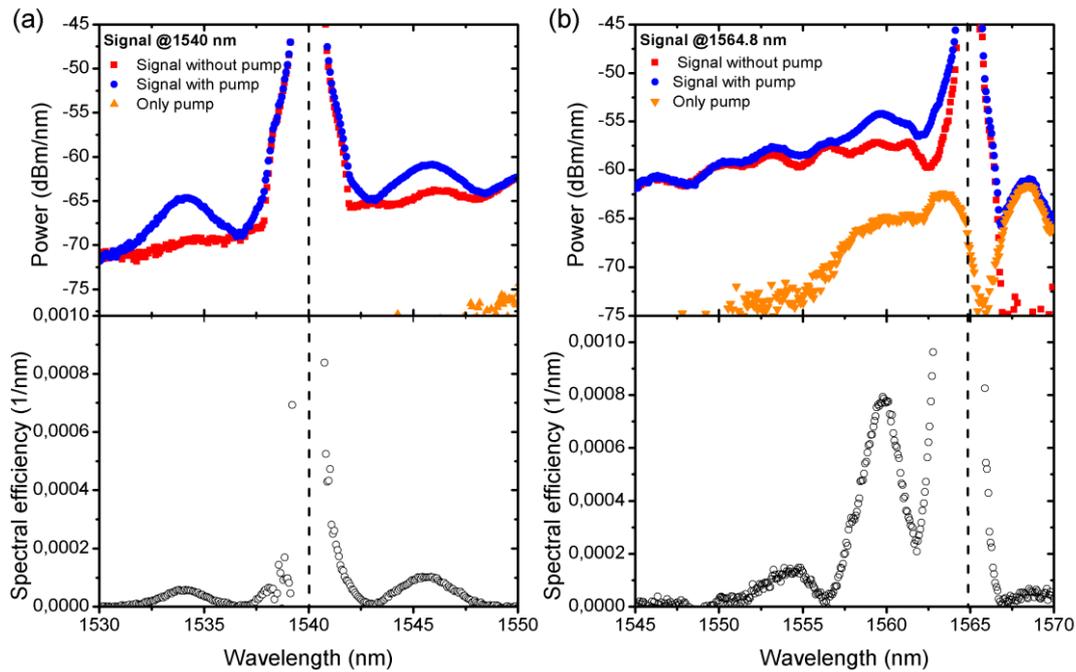

**Figure S4**: Experimental spectra recorded at the output of our silicon Bragg grating waveguide of an input signal at a wavelength of (a) 1540 nm and (b) 1564.8 nm and pump peak power of 15 W. The spectrum of the CW signal that has not interacted with the pump pulses (red trace), of the pump pulses alone (orange trace), and of both the pump pulses and the signal (blue trace) are shown on a logarithmic scale (upper graph). They are subtracted on a linear scale (black trace) in the lower graph, in order to obtain only the converted signal without the residual of the CW input signal. The dashed line represents the center wavelengths of the input signals light.

**Supplementary Note 4: Free carrier generation and index change of an optical mode**. The time dependence of the free carrier (FC) density $N(t)$ at a fixed point in space generated by two-photon absorption (TPA) of a pump pulse propagating in a silicon waveguide is described by [26]:

$$\frac{dN(t)}{dt} = \frac{\beta}{2hf}\left(\frac{n_g}{n_{si}}\right)^2 I^2(t) - \frac{N(t)}{\tau_{FC}} \tag{S1}$$

The first term on the right-hand-side of this equation describes the generation of free carriers via the TPA process, while the second term describe the decay of the generated free carriers, characterized by the lifetime $\tau_{FC}$. Here, $h$ is Planck's constant, $\beta = 0.5$ cm/GW is the TPA coefficient in silicon at a frequency $f = 193.4$ THz [26], $n_g$ is the pump group index, $n_{si}$ is the silicon refractive index, and $I(t)$ is the pump pulse intensity.

The FC induced refractive index change can be described by the empirical relation presented by Soref et al. [42]:

$$\Delta n_{FC}(t) = -8.8 \cdot 10^{-22} N(t) - 8.5 \cdot 10^{-18} N(t)^{0.8} \tag{S2}$$

While the Kerr induced refractive index change is described by:

$$\Delta n_{Kerr}(t) = n_2 \cdot I(t) \tag{S3}$$

where $n_2$ is the nonlinear refractive index. Therefore, the total induced index change at the input of the waveguide can be described by:

$$\Delta n_{Total}(t) = \Delta n_{FC}(t) + \Delta n_{Kerr}(t) \tag{S4}$$

We integrate the Equation S1 numerically assuming a Gaussian pulse intensity $I(t) = I_0 \exp\left(-4ln2 t^2/t_p^2\right)$, where $t_p$ is the pump pulse duration, and neglecting the FC recombination term as FCs have long life time in comparison to pulse duration.

The maximum of the generated FC density is given by:

$$N_{FC,max} = \frac{\beta}{2hf}\left(\frac{n_g}{n_{si}}\right)^2 I^2 \sqrt{\pi}\frac{t_p}{\sqrt{8ln2}} \tag{S5}$$

When a small perturbation of the refractive index is applied, the frequency of an eigenmode will change according to [43]:

$$\Delta\omega = -\frac{\omega}{2}\frac{\int \Delta\epsilon(\vec{r})|\vec{E}(\vec{r})|^2 d\vec{r}}{\int \epsilon(\vec{r})|\vec{E}(\vec{r})|^2 d\vec{r}} \qquad \text{(S6)}$$

In our case most of the energy is guided in the waveguide and that is where the FC are generated. Thus we simplify the Equation S5 with the relation:

$$\frac{\Delta\omega}{\omega} = -\frac{\Delta n}{n} \qquad \text{(S7)}$$

From this equation we can estimate the expected frequency shift of the optical mode according to the maximum refractive index change $\Delta n$.

For $t_p = 2$ ps, $n_g = 3.5$, a pulse peak power of $\approx 15$ W. at waveguide input, and by taking into account an effective mode area of the pump pulse of $0.1\ \mu m^2$ [44] and $n_2 = 9 \cdot 10^{-14}\ cm^2/W$, we calculate the temporal index change at the input of our waveguide using Equation S4. Here the maximum generated FC density is $\approx 9.5 \cdot 10^{17}\ 1/cm^3$. The simulation results are presented in **Figure S5**. Orange, red, and black solid lines represent the index change due to free carriers injections $\Delta n_{FC}$, Kerr effect $\Delta n_{Kerr}$, and the total index change $\Delta n_{Total}$, respectively. We define the slope $\partial \Delta n/\partial t$ as the slope at a point of $\partial^2 \Delta n/\partial t^2 = 0$ (green dashed line in **Figure S5**), which is in this case equal to $\Delta n_{eff}/\tau$, where $\tau$ is the effective rise time of the negative front. $\tau$ is defined as the time from the maximum of the total index change $\Delta n_{Total}$ till the crossing point of the green dashed line with the saturation line of $\Delta n_{Total}$ (grey circles in **Figure S5**). According to Equation S7, this index change $\Delta n_{eff}$ will lead to a wavelength change $\Delta\lambda_D$, i.e. $\partial \Delta\lambda_D/\partial t = \Delta\lambda_D/\tau$. Therefore, the maximum possible wavelength shift of the signal in the structure $\Delta\lambda = \Delta\lambda_D \cdot \Delta t/\tau$, can be estimated by assuming that the maximal time spent inside the front $\Delta t$ is equal to the pump travelling time inside the 1 mm structure which is $\approx 12$ ps, $\Delta\lambda_D = 2.2$ nm, and $\tau = 1.9$ ps.

This yields to maximum wavelength shift of ≈ 13 nm, which approximately fits to the maximal obtained experimental shift of ≈ −11 nm in the experiment.

We should also mention, that the pump power decays in the waveguide due to absorption and scattering, and accordingly does the $\Delta n$.

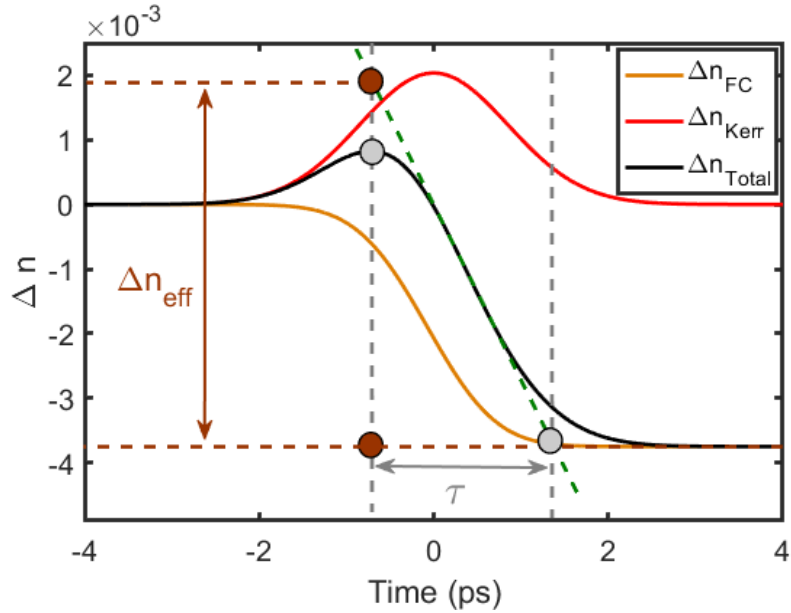

**Figure S5**: Simulation of the induced refractive index change as a function of time by a 2 ps-long pump pulse propagating in the silicon waveguide. Orange, red, and black solid lines represent the index change due to free carriers injections $\Delta n_{FC}$, Kerr effect $\Delta n_{Kerr}$, and the total index change $\Delta n_{Total}$, respectively. We define the slope $\partial \Delta n / \partial t$ as the slope at a point of $\partial^2 \Delta n / \partial t^2 = 0$ (green dashed line), which is in this case equal to $\Delta n_{eff} / \tau$, where $\tau$ is the effective rise time of the negative front.